# DC-SQUID Sensitivity for Readout of Entangled State Quantum Bits

H. Tanaka, Y. Sekine, S. Saito and H. Takayanagi
NTT Basic Research Laboratories
Morinosato-Wakamiya 3-1, Atsugi, Kanagawa, 243-0198, JAPAN

We propose a scheme for entangled state measurement at flux qubits (quantum bits) depending on the sensitivity measurement of a DC-SQUID (Superconducting QUantum Interference Device). The DC-SQUID is used as a flux qubit readout device. The switching current distribution of a DC-SQUID is sufficiently narrow to distinguish between two ground states of three Josephson junction qubits with a large mutual inductance between a qubit and the DC-SQUID. But discrimination between two ground states becomes difficult with the smaller mutual inductance that is preferable as regards of qubit coherence. However, we can employ averaging to increase the effective sensitivity. And we found that Bell-pair measurement can be performed with flux qubits to prove the existence of an entangled state over two qubits.

## I. INTRODUCTION

Quantum computation is a novel architecture that achieves extremely fast computation. The basic element of a quantum computer is a qubit, which is expressed as a quantum two level system in quantum physics. Several two level physical systems are thought to be qubit candidates. In principle, any quantum two level system has the potential to be applied to qubits. However each system has its own advantages and disadvantages. Important criteria as regards qubit feasibility are tolerance to decoherence, efficient qubit interaction and scalability. Although solid-state quantum computers must be improved in terms of coherence, qubits in a solid state such as a semiconductor or superconductor have appropriate scalability when we utilize the well-established nanometer scale fabrication technology now widely used in the semiconductor industry. The short coherence time in a solid-state quantum bit is due mainly to the existence of many degrees of freedom as found with electrons and phonons in a solid state. Gate operations in a solid-state quantum computer have not been demonstrated experimentally yet. In contrast, the gate operations have already been demonstrated in NMR[1] and ion trap quantum computers, for example controlled NOT. NMR has even been used to perform practical algorithms, such as Deutsch-Jozsa, quantum search and Shor's factorization[2], with several qubits. But experimental schemes in NMR are not directly applicable to quantum computers over 10 qubits because of the short coherence time and initialization problem.

Quantum computing operations more than 2 qubits always require an entangled state. We must control and maintain entangled state during gate operation. Entanglement is usually used to obtain a quantum correlation between two particles, such as phonons or electrons. A photon pair of spin singlet state is one type of entangled state. Two particles in a spin singlet state, which is sometimes called a Bell-pair state, is a maximally entangled state. Realization of universal gate operations requires full controllability of any kind of entangled state. Furthermore, entangled state readout is also needed. Therefore it is important to realize an entangled state over two quantum bits as regards quantum computation and temporal coherence control of a wavefunction in a coupled quantum system in a solid state, in which a large number of particles are involved in the dynamics.

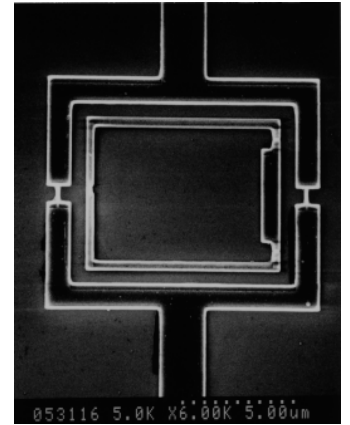

Fig.1 SEM micrograph of qubit (inside) and DC-SQUID (outside). Three constricted regions in inside qubit and two in outside DC-SQUID are Josephson junctions.

## II. FLUX QUBITS

With solid-state quantum computers, a superconductor has a longer coherence time because of the superconducting energy gap. This is the energy difference between the condensed Cooper-pair state and the normal state. This energy gap protects our qubits from unexpected excitations that probably originate with the environment and destroy qubit coherence, leading to errors or information loss during computation.

Both charge and flux qubits have already been confirmed experimentally. The charge qubit[3] has already realized full one-qubit operation, but it has a large charge noise in its Josephson junctions. This charge noise, which sometimes shows 1/f characteristics, is unavoidable when we use an evaporated metal superconductor with the Dolan[4] bridge technique, which is the only method currently available for fabricating these types of junctions smaller than 100 nm. In contrast, we employ the same material and technique for flux qubits but they are insensitive to charge noise. Therefore we can expect a longer coherence time than with a charge qubit.

We are studying three Josephson junction flux qubits that were first proposed by Mooij et al.[5] Figure 1 shows this qubit, located in a DC-SQUID. The DC-SQUID is a highly sensitive device for magnetic field measurement. Both the qubit and DC-SQUID in Fig.1 were fabricated on a thermally oxidized silicon wafer with aluminum and aluminum oxide as a superconductor and an insulator, respectively. The silicon oxide was 2μm thick. The double evaporated aluminum layer was 50nm thick. The square loop inside is a superconductor ring with three junctions and has a persistent circulating current. This current generates flux through the ring and the flux orientation is determined by the current direction. The magnitude of the flux is quantized due to fluxoid quantization along the superconductor ring. The two lowest states, ground and the first excited state, are used to store quantum information as |0> and |1>. These two states also correspond to clockwise and counterclockwise current. The direction of generated flux from qubit is determined by current direction.

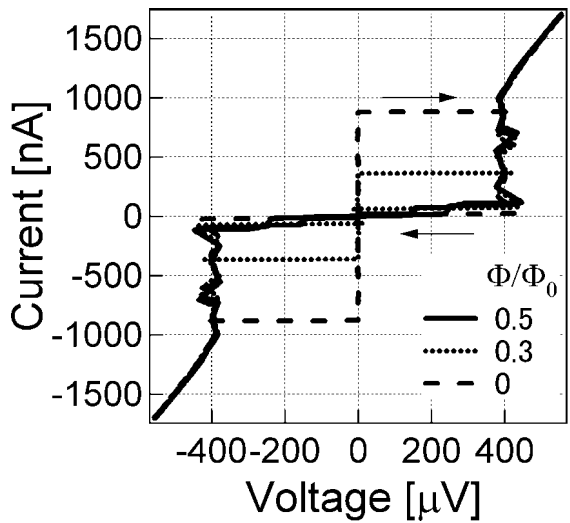

Fig.2 Current voltage characteristics of DC-SQUID. This DC-SQUID has no inside qubit for native DC-SQUID characterization. Current is swept both in positive and negative direction. Current direction is guided by arrow whose filling is 0. External magnetic field filling against flux quantum is fixed at 0, 0.3 and 0.5, respectively.

Flux qubit states can be read out by measuring the magnetic field with an under-damped DC-SQUID, namely without a shunt resistor. Although a standard DC-SQUID has shunt resistors to prevent the hystereses in the current-voltage characteristics, we used an un-shunted SQUID to minimize the coupling between the qubit and the dissipative environment introduced by the shunt resistors. It is important that we perform effective measurement without disturbing the qubit state during operation. These two requirements are somewhat contradictory. Effective measurement is performed by increasing the mutual inductance between a qubit and the DC-SQUID. We can control the mutual inductance by changing the distance between qubit and DC-SQUID. Large inductance provides good sensitivity but leads to large decoherence.

## III. ONE QUBIT READOUT

Figure 2 shows typical current voltage characteristics for a DC-SQUID. DC-SQUID measurement is current biased and the switching current is measured where the voltage across the DC-SQUID jumps from zero to a finite voltage state. Large hysteresis is observed in Fig. 2. This indicates the fact that we used an un-shunted DC-SQUID. The switching current depends on the magnetic flux penetrating the DC-SQUID and exhibits the periodical behavior shown in Fig. 3. The horizontal axis in Fig. 3 is an external magnetic field. There is a small discontinuous region where the ground state of the inner qubit changes and the orientation of the generated flux switches. The qubit generated magnetic flux picked up by the DC-SQUID is smaller than the flux quantum. The measured magnetic flux is the sum of external and qubit generated magnetic flux. The global periodical dependence in Fig. 3 is caused by an external magnetic field and the discontinuous step is caused by an inside qubit. Figure 4 is an enlarged graph of the discontinuous region in Fig. 3. The plot consists of small points. Each point describes a single measurement. A current sweep provides a single measurement at its switching to a finite voltage state. Figure 4 corresponds to around 5000 measurements. We clearly see two lines that correspond to clockwise and counterclockwise qubit current. This separation means that the DC-SQUID can detect changes in qubit ground states although these two states are both in the ground states. We did not apply any excitation microwave to the qubit in this measurement. An excitation microwave causes a transition between the ground and excited states. Either state has to be in the first excited state for real qubit operation. However the flux change magnitude is the same in our experiment. Therefore we can estimate the readout device feasibility by considering the sensitivity of this measurement result.

Switching happens on a lower line when the filling is reasonably smaller than 0.5. The probability of finding a switching current at the lower and upper lines is even the same at a filling of 0.5.

We closely investigated the sensitivity of the readout DC-SQUID. The relative width of the switching current distribution is $\Delta I_{SW}/I_{SW} \cong 0.002$ where $I_{SW}$ and $\Delta I_{SW}$ are the maximum switching current of the DC-SQUID and the standard deviation of the switching current, respectively. The maximum switching current can be read out from Fig. 3 and is $I_{SW} \cong 165 nA$. The switching current difference due to the qubit state is $\Delta I_{QUBIT} \cong 5 nA$. The ratio between $\Delta I_{QUBIT}$ and $\Delta I_{SW}$ is $\Delta I_{SW}/\Delta I_{QUBIT} \cong 0.064$. This means that discrimination is sufficient for qubit readout.

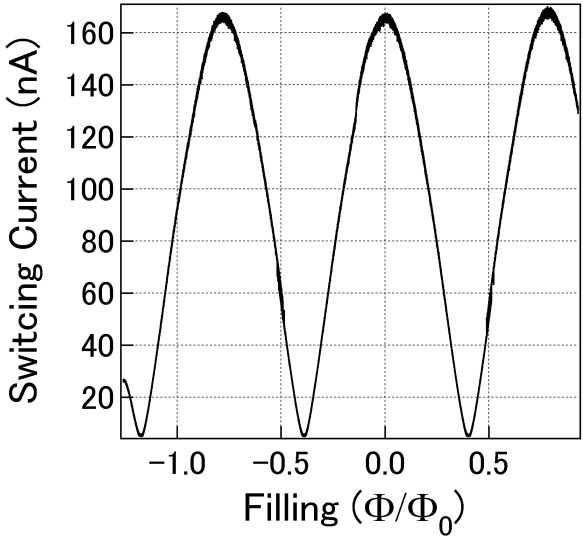

Fig.3 The switching current dependence versus filling factor of external magnetic flux through qubit.

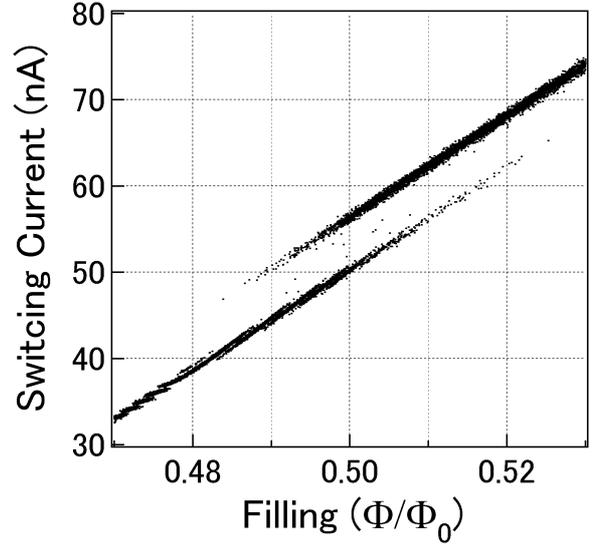

Fig.4 Enlarged graph of Fig.2. Upper and lower lines are two ground states of quit and have about 5 nA difference.

We designed this qubit and DC-SQUID to obtain a relatively large output signal. We achieved this by designing the qubit and DC-SQIUD with a large mutual inductance between them. But a large mutual inductance is not appropriate for a qubit. A qubit with a reasonable coherence time will give us a smaller readout signal for the time being, as reported by Casper et al.[6] Figure 5 shows the switching current dependence for the sample with a relatively small mutual inductance. This DC-SQUID has a larger maximum switching current $I_{SW} \cong 340 nA$. But the relative sensitivity is $\Delta I_{SW}/I_{SW} \cong 0.002$, which is almost the same as that described previously. Absolute magnetic flux sensitivity depends on $\Delta I_{SW}/I_{SW}$. Although we can still read out the switching current difference of the qubit state, the two lines are closer together in Fig. 5. The switching current change induced by the qubit is $\Delta I_{QUBIT} \cong 2.5 nA$, which is almost half that in Fig. 4. This decrease is result from the change in the mutual inductance. The ratio between $\Delta I_{QUBIT}$ and $\Delta I_{SW}$ is $\Delta I_{SW}/\Delta I_{QUBIT} \cong 0.27$. This number is much larger than that in the previous sample. This can also be described by looking at Fig. 4 and Fig. 5 and comparing the scattering width of the plot points against the gap between the lower and upper lines. In Fig. 5, the scattering width and gap are almost comparable even though the factor, 0.27, is somewhat smaller than unity. This implies that we are taking the standard deviation as a switching current distribution as $\Delta I_{SW}$. What we see at the plot points as a width of the scattering region is fairly close to the peak-to-peak value. We are assuming that the switching current distribution $\Delta I_{SW}$ has a normal distribution, which is almost the same as the experimental data except for a slight difference in the lower current side of the distribution.

In this region of mutual inductance, we are unable to distinguish a qubit state in a single measurement. But we can distinguish qubit states by averaging thousands of measurements. We realize this by counting the frequency of SQUID switching events at a certain current. Each current sweep is made independently, therefore this scheme provides a time ensemble average of independent events. Based on this assumption, $N$ measurements and averaging give us a $\sqrt{N}$ times smaller standard deviation. Then the effective standard deviation $\Delta I_{SWAVE}$ become $\Delta I_{SWAVE} = \Delta I_{SW}/\sqrt{N}$. Finally the factor that determine the discrimination ability become $\Delta I_{SWAVE}/\Delta I_{QUBIT} = \Delta I_{SW}/\sqrt{N}\Delta I_{QUBIT}$. This means that we can enhance the effective sensitivity simply by increasing the number of measurement.

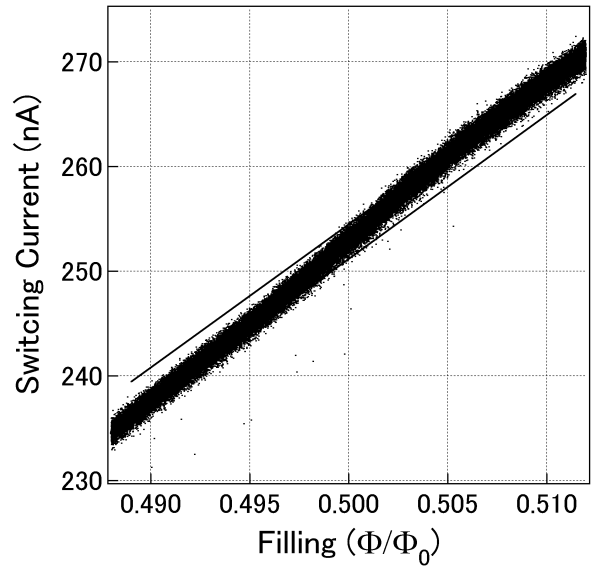

Fig.5 Enlarged graph of switching current dependence for smaller mutual inductance between qubit and DC-SQUID.

IV. ENSEMBLE AVERAGE and ENTANGLEMENT.

In essence the speed of a quantum computer involves quantum parallelism based on superposition and entanglement. Of course quantum computation is not deterministic during its operation but certain algorithm such as Deutsch-Jozsa gives us completely deterministic answers that are yes-or-no results of function characteristics. But Shor's algorithm, a quantum Fourier transformation, sometimes gives superpositional states as its answer before the measurement has been performed. An ensemble averaged readout is not efficient for this kind of algorithm and sometimes does not work. A similar situation occurs if we apply a quantum computer at the front end of a quantum cryptography system. This would require the ability to handle a true one qubit readout with a single measurement, rather than an ensemble average. NMR quantum computer has already demonstrated all three of the above algorithms, except for cryptography application. But the readout process of an NMR quantum computer is completely ensemble averaged over Avogadro's number of molecules. This type of quantum computer is sometimes called an expectational value quantum computer. The superconductor charge qubits readout is also ensemble averaged. We can know only the averaged value of the qubit state over thousands or millions of events.

The point is whether ensemble averaging is performed before or after the physical measurement. "Pre-ensemble measurement" is a scheme where the measurement is performed before the averaging. With "post-ensemble measurement" the measurement is performed after the averaging. Usually with "post-ensemble measurement", the averaging scheme is incorporated into the physical measurement process almost automatically. For example, a pick up coil in an NMR can only detect the collective signals from all the molecules therein. It is impossible for us with available technology to detect the signal from one molecule. In a superconductor charge qubit, the readout is measured by the current which is also the collective physical quantity from millions of excitation events. We cannot currently detect the charge of a single Cooper-pair. In contrast, in the "pre-ensemble measurement", we make multiple measurements and then calculate the average value explicitly. No pre-ensemble measurement quantum computer has been experimentally realized but a flux qubit can be a "pre-ensemble measurement" qubit. We should also note that the qubits with a single photon using cavity QED (Quantum ElectroDynamics) with a polarizer and single photon counting, which are also being intensively studied, can be "pre-ensemble measurement" qubits. With flux qubits with an un-shunted DC-SQUID, we can know the readout values of a single event even if the sensitivity is insufficient to distinguish two states in one measurement. But flux qubits with continuous measurement using a shunted DC-SQUID[7] do not constitute a "pre-ensemble measurement" because the measured value has already been averaged in the same as with charge qubits.

A significant advantage of "pre-ensemble measurement" is that the entanglement of two qubits remains as a correlation between the readout values of each qubit. Therefore we can confirm the existence of entanglement in flux qubits by measuring the switching current. We can also say that it is possible to perform Bell-pair like experiments in a quantum computer for "pre-ensemble measurement". One way to prove the existence of entanglement is to show a quantum correlation between two qubits. And a Bell-pair measurement experiment can prove the entanglement. This is achieved by taking the correlation of two qubits over multiple measurements and inserting it into Bell's inequality[8].

In Bell's inequality we have to perform a local measurement for two qubits with arbitrary angles. Our DC-SQUID can read only generated flux which is directly coupled to the phase in Josephson junctions. The phase in a Josephson junction is a conjugated value against the charge number at the capacitance across the junctions. These two variables, phase and charge, are non-commuting variables in quantum physics like the position $x$ and momentum $p$ of a particle. The dynamics of a qubit is expressed by a quantum two level system which is mathematically the same as the spin 1/2 system. Therefore we can express the qubit state with pseudo spin. Suppose we choose bases of a flux qubit as $\sigma_Z$, then the DC-SQUID measures along the Z-direction of the pseudo spin. Another angle measurement such as $\sigma_X$, $\sigma_Y$ or the angle in between is performed as follows. The first step is to create a spin rotation with a pulse and the next is to measure the pseudo spin along the Z-direction. DC-SQUID can measure only Z-direction of pseudo spin. However previous rotation makes it possible to measure along an arbitrary pseudo spin angle by choosing an appropriate phase and amplitude duration product for the pulse. Employing Z-direction measurement and arbitrary angle rotation, we can measure any angle. By applying this rotation pulse to either qubits we can perform local measurement on either side of the Bell pair.

V. SIMULATION.

We performed a simulation to confirm the feasibility of Bell pair measurement for flux qubits based on the experimental switching current distribution result because we were unable to determine the qubit state with a single measurement under large distribution. We show the simulation result for the simplest condition even though we can only see a classical correlation from this situation. Considering the independence of each measurement, we can extend this result to the general Bell's inequality. Then we are able to confirm the existence of quantum correlation by choosing the measurement angles required by Bell's inequality.

Figure 6(a) and (b) show simulated results of two qubit measurements for flux qubits with different switching current distributions for maximally entangled states, $(|01>+|10>)/\sqrt{2}$. We applied simple Monte Carlo methods to two qubits with ideal interaction. Each state is completely separated from the other in Fig. 6(a) with $\Delta I_{SW}/\Delta I_{QUBIT}=0.04$ where $\sigma$ is the

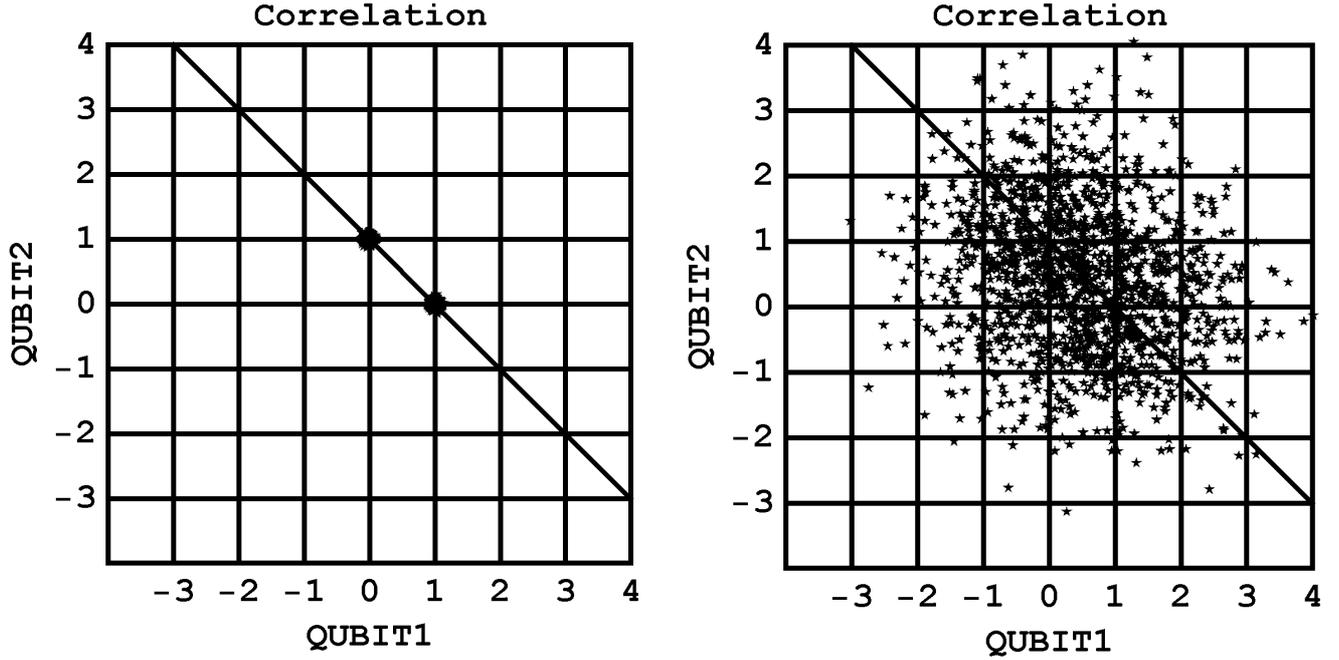

Fig.6 (a), left side, Monte Carlo simulated readout values of each qubit in entangled state with $\Delta I_{SW}/\Delta I_{QUBIT}=0.04$ and (b), right side, with $\Delta I_{SW}/\Delta I_{QUBIT}=1.0$. Horizontal and vertical axes correspond to the readout values of qubit1 and qubit2, respectively. |01> and |10> happen with even probability with ideal case that means the standard deviation is 0.

normalized standard deviation. The two axes indicate the normalized readout value of qubit1 and qubit2, respectively. The two location points correspond to the states, |01> and |10>. Each of these states is read out with 50% probability. In contrast, the readout values of the two states overlap each other in Fig. 6(b) with $\Delta I_{SW}/\Delta I_{QUBIT}=1.0$. This value is even larger than that in Fig. 5 that has $\Delta I_{SW}/\Delta I_{QUBIT}\cong 0.27$. But if we take the cross correlation between the readout values of the two qubits, it clearly shows the correlation brought about by the entanglement. The solid line shows this correlation. Therefore, it is possible to confirm experimentally the entanglement over two qubits by adopting the cross correlation even with a relatively large switching current distribution.

V. SUMMARY.

We showed an experimental DC-SQUID readout result and found that the DC-SQUID has good sensitivity. However to obtain a readout with better qubit coherence, we have to employ an ensemble averaging scheme due to the small mutual inductance. We used a simulation to show that the entanglement of flux qubits could be confirmed with this DC-SQUID using Bell's inequality with an ensemble averaged value.

ACKNOWLEDGMENT

We thank J. E. Mooij and C. J. P. M. Harmans for motivating this research, and Caspar H. van der Wal and Y. Nakamura for useful discussions.